# FRACTIONAL CHARGE DETERMINATION VIA QUANTUM SHOT NOISE MEASUREMENTS

M. Heiblum

*Braun Center for Submicron Research, Department of Condensed Matter Physics,*
*Weizmann Institute of Science, Rehovot, 76100, Israel*

Charge excitations in a two dimensional electron gas, under a quantizing magnetic field and in the fractional quantum Hall effect regime, flow in one dimensional-like strips along the edges of the sample. These excitations (quasiparticles) may be independent or condense into an interacting chiral Luttinger liquid. Adding a backscattering potential, which reflects a forward propagating quasiparticle to a backward propagating one, partitions the stream of quasiparticles and induces quantum shot noise. The noise is proportional to quasiparticles charge and may be affected by their mutual interactions. The dependence of the determined charge on the temperature, excitation energy, and partitioning will be describes for a few fractional states, revealing in some cases a *universal* behavior.

## 1. Introduction

Shot noise refers to time-dependent fluctuations in an electrical current, being a direct consequence of the *particle*-like nature of the electrons and the stochastic nature of their injection. Shot noise generally provides information on the charge and the mutual correlations of the particles - not necessarily given by the time-averaged current. The average current, according to Landauer formula, merely probes the average transmission coefficient of the electrons through the conductor. In a truly classical stochastic (Poisson-like) electron emission processes the time average, or the ensemble average, of the squared current fluctuations, is $\langle i^2 \rangle_{\Delta f} = S_i(f) \cdot \Delta f = 2eI_t \cdot \Delta f$, where $S_i(f)$ is the generally *white* spectral density, $I_t$ the average transmitted current, and $\Delta f$ the bandwidth of the measurement. In mesoscopic conductors, however, where electrons scatter only elastically, shot noise is suppressed. The reason for this suppression is fundamental, resulting from the fermionic nature of the electrons [2]. In contrast, in classical conductors, where electrons scatter many times and thermalize, shot noise, in the sense of being proportional to the current, is not observed, with the noise expressed via the increased temperature and thus the thermal noise.

We are concerned here with quantum Hall effects, observed in a two dimensional electron gas (2DEG) under a strong magnetic field $B$. The energy spectrum of a 2DEG is then consisting of highly degenerate Landau levels with a degeneracy per unit area $d=B/\phi_0$, with $\phi_0=h/e$ the flux quantum ($h$ being Plank's constant). Whenever the magnetic field is such that an integer number $\nu$ (the *filling factor)* of Landau levels are occupied, that is $\nu=n_s/d$ equals an integer ($n_s$ being the areal density of the 2DEG), the longitudinal conductance of the 2DEG vanishes while the Hall conductance equals $\nu e^2/h$ with very high accuracy. This is the celebrated *integer quantum Hall* effect (IQHE) [3,4]. A similar behavior of the conductance is found at *fractional* filling factors, namely, when the filling factor equals a rational fraction with (mostly) an odd denominator $2p+1$, being known as the *fractional quantum Hall* effect (FQHE) [4,5]. In contrast to the IQHE, which is well understood in terms of non-interacting electrons, the FQHE cannot be explained in such terms and is believed to result from interactions among the electrons, brought about by the strong magnetic field. Laughlin's theoretical prediction of the existence of fractionally charged *quasiparticles*, each having a fractional charge $e^*=e/(2p+1)$, e.g., $e/3$, $e/5$ and $e/7$, put forward in order to explain the FQHE effect [6], is very counterintuitive.

Halperin was the first to suggest edge channels transport to explain the conduction mechanism in the IQHE [7]. According to this successful model current flows along the edges of the sample, with electrons performing classical chiral 'skipping orbits'. In a quantum mechanical language, current flows at the crossing of the Landau levels with the chemical potential, near the edges of the sample. In the IQHE regime, electron-correlations are weak and currents are presumed to flow in non-interacting 1d like channels. Alternatively, in the FQHE regime, electron correlations are strong, leading the edge channels to behave as 1d chiral Luttinger liquid [8]. Due to the chirality, backscattering in wide samples is minimized and edge channels propagate void of backscattering for long distances.

Edge channels produce shot noise (called sometimes 'excess noise') on top of the ubiquitous thermal noise ('Johnson-Nyquist' noise), with the latter being a property of any conductor and is independent of its microscopic properties and the electron charge. Its spectral density $S_T=4k_BTg$, with $k_B$ the Boltzman's constant, $T$ the temperature, and $g$ the conductance, originates from microscopic current fluctuations (with no current flowing) due to the finite temperature of the electrons. The so called quantum shot noise differs from the classical one reflecting the noise free property of the emitting reservoir (due to its fermionic nature) [2,9,10]. This was first demonstrated in the simplest mesoscopic system: a ballistic constriction in the 2DEG. The constriction was formed by Quantum Point Contact (QPC) [11], being two closely separated metallic gates evaporated on the surface of the heterostructure embedding the 2DEG (see inset in Fig. 1). At zero temperature the contribution to the shot noise of the $p$'th propagating channel in the constriction is:

$$S_i = 2e^*Vg_p t_p(1-t_p) \quad , \tag{1}$$

where $S_i$ stands for the low frequency ($f<<eV/h$) spectral density of current fluctuations, $V$ the applied source-drain voltage, $g_p$ the conductance of the fully transmitted $p$'th channel, and $t_p$ the effective transmission coefficient of the $p$'th channel. This reduces to the well known *classical* poissonian expression for shot noise (above) when $t_p<<1$ ('Schottky equation').

## 2. Noise of Fractional Edge Channels

Recent theoretical studies of shot noise in the FQHE regime, based on the chiral Luttinger liquid model, are applicable only to the *Laughlin fractional states*, e.g., $v=1/3, 1/5,$ etc. [8], namely, when the edge current is carried by only one channel. They predict a spectral density of shot noise very much like in Eq. 1, with two limits [12]:

$$\begin{aligned} S_{T=0} &= 2e^*Vg_p(1-t_p) = 2e^*I_r \quad ; \quad t_p \approx 1 \\ S_{T=0} &= 2eVg_p t_p = 2eI_t \quad ; \quad t_p \approx 0 \end{aligned} \tag{2}$$

where $I_r$ and $I_t$ are the reflected and transmitted average currents, respectively. The most important outcome of Eq. 2 is the prediction that the tunnelling of quasiparticles $e^*$ contributes to the shot noise at weak reflection ($t_p \approx 1$). No easily accessible formulation for an arbitrary transmission exists [12].

One can gain insight into the characteristics of the expected shot noise in the FQHE regime by considering the composite fermions (CF) model [13]. In its simplest form the model identifies the fractionally filled first electron Landau level, $\nu=p/(2p+1)$, as an integer $p$ filled Landau levels of CFs ($\nu_{CF}=p$). Each CF consists of an electron and *two* magnetic flux quanta opposing the original magnetic field attached to it. Hence, the net magnetic field sensed by the CFs is $B-2n_sh/e$ (equals to zero at $\nu=1/2$). Under this weaker effective magnetic field the CFs can be approximated by weakly interacting quasiparticles, flowing in separate and nearly non-interacting edge channels, hence, justifying the application of the above mentioned formulae for stochastic back scattering of particle and a poissonian shot noise. When the QPC partly pinches off and the conductance is in a transition between two different FQHE plateaus of the series $p/(2p+1)$, only the highest edge channel is being partitioned; the other, lower lying channels, are fully transmitted. Consequently, in Eqs. (1) and (2), $p$ designates the CF edge channel that is being partitioned. If in the transition region between $\nu=1/3$ and an insulator the values are $p=1$, $g_1=g_Q/3$ and $t_1=3g/g_Q$, then, in the transition region between $\nu=2/5$ and $\nu=1/3$ they are $p=2$, $g_2=(2/5-1/3)g_Q$, and $t_2 = \dfrac{g/g_Q - 1/3}{2/5 - 1/3}$. Here $g$ is the total two-terminal conductance and $g_Q=e^2/h$ the quantum conductance.

A more general expression for shot noise of non-interacting particles [14], applicable at finite temperatures, is:

$$S = 2e^*V\, g_p\, t_p(1-t_p)\left[\coth(\dfrac{e^*V}{2k_BT}) - \dfrac{2k_BT}{e^*V}\right] + 4k_BTg \quad , \quad (3)$$

with a finite noise at zero applied voltage, $S_T=4k_BTg$. When $V>V_T \sim 2k_BT/e^*$ the noise approaches the linear behavior predicted by Eqs. 1 and 2. This expression was utilized for charge determination over a wide range of conditions (transmission and temperature); as described later. It should be noted that a different approach to determine charge, e.g., based on visualization of localized quasiparticles [15] and resonant tunnelling of quasiparticles into an isolated island (or an impurity) [16].

## 3. Experimental Considerations

In order to realize shot noise measurements we formed constrictions in the plane of a 2DEG, embedded in the heterostructure, some 100nm below the surface. Carrier densities are around $n_s=(1-2)\cdot 10^{11}\text{cm}^{-2}$ and nobilities $\mu=(2-30)\cdot 10^{6}\text{cm}^{2}/\text{V-s}$ at 4.2K. Samples were cooled in a dilution refrigerator to an electron temperature 10-50mK. Current fluctuations, generated in the constriction, were fed into a *LC* resonant circuit with center frequency ~1MHz and bandwidth of 30-100kHz. The fluctuations were amplified by a nearby extremely low noise, home made, preamplifier, cooled to 4.2K. The preamplifier has a voltage noise as low as 2.5 $10^{-19}\text{V}^{2}/\text{Hz}$ and a current noise of some $(4-10)\cdot 10^{-29}\text{A}^{2}/\text{Hz}$. Outside the cryostat the amplified signal was fed into an additional amplifier followed by a spectrum analyzer. Since the accurate magnitude of the noise signal is important, a careful calibration of the total gain is routinely done by utilizing a calibrated current noise source. Measuring the thermal noise generated by the conduction of the constriction (or by the quantum conductance in a multi-terminal configuration) as function of the inverse conductance $1/g$, both the electron temperature, via $T=(\delta S_T/\delta g)/4k_B$, and the current noise of the amplifier (extracted from the extrapolated total noise to zero inverse conductance) were deduced (as shown in Fig. 1).

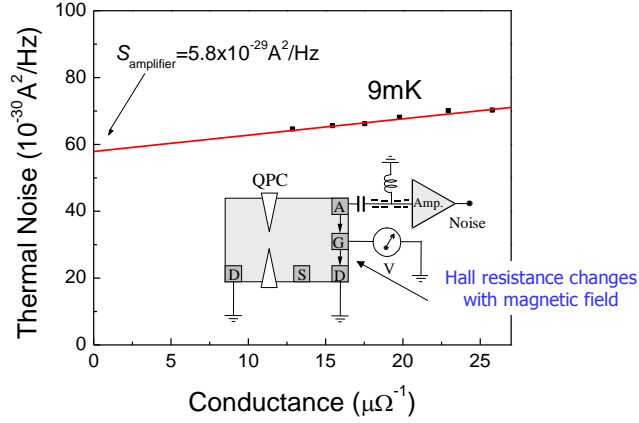

**Fig. 1.** A general configuration of the noise measurement system with the measured shot noise at equilibrium at the input of the preamplifier as a function of the output conductance. The output conductance is independent of the transmission of the QPC and is controlled via the magnetic field. The measured noise is a sum of thermal noise $4k_BTg$ and the constant noise of the amplifier.

## 4. Shot Noise at $\nu=1/3$

Shot noise measurements as a function of the current through the constriction were performed first in the absence of magnetic field. The results, after calibration and subtraction of the amplifier's noise, are shown in Fig. 2. The transmission of the lowest laying quasi 1d channel was simply deduced from the measured conductance normalized by $2e^2/h$ (the factor 2 accounts for spin degeneracy). Our data almost perfectly fits the expected noise from electrons predicted by Eq. 2 using the independently measured electron temperature [17].

At high magnetic field (10-14Tesla), the two-terminal conductance exhibits Hall plateaus at fractional fillings. We start by exploring filling factor $\nu=1/3$ with $g_{1/3}=e^2/3h$. Figures 3a and 3b present typical differential conductance of the constriction as function of the applied voltage at electron temperatures $T=23$mK and $T=120$mK, for different backscattering potential strengths (determined by the split-gate voltage of the QPC). At the lower temperature, even a relatively weak backscattering potential (transmission $t=g/g_{1/3}\sim 0.7$ at high bias) almost fully reflects the current at zero applied voltage.

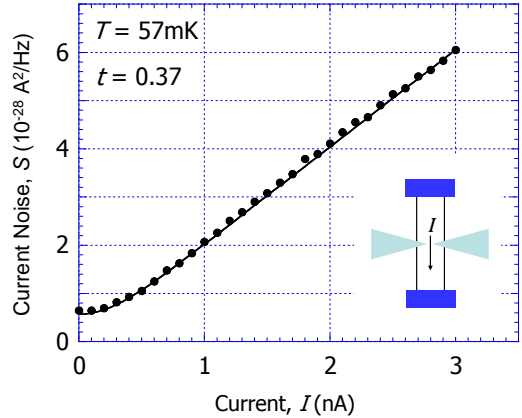

**Fig. 2.** Shot noise as a function of the current $I$ through the constriction without an applied magnetic field. The solid line is the noise given by Eq. 3 with the temperature deduced independently (as in Fig. 1).

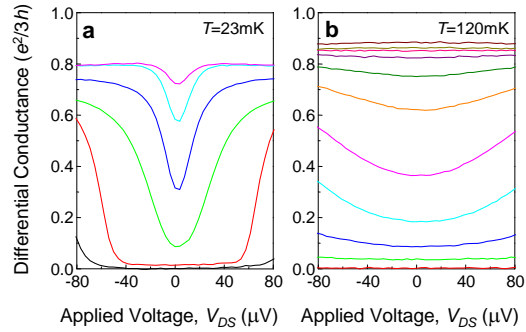

**Fig. 3.** The differential conductance of the constriction at $\nu=1/3$ as function of $V_{DS}$. **a.** Measured at 23mK for a few backscattering potential strengths; from bottom to top: QPC split-gate voltage changes from -83mV to -0.5mV. **b.** Similar data measured at 120mK; from bottom to top: QPC split-gate voltage changes from -131mV to -11mV.

We turn now to the shot noise in this regime. The conductance and shot noise at different temperatures are plotted as function of the applied voltage across the constriction in Fig. 4. At 23mK the conductance is highly non-linear and the noise is non-poissonian. The noise has only the qualitative features of the CLL, predicted by Fendley et al. [18]. The quantitative disagreement may be related to a deviation of the scattering potential and its voltage dependence from the assumed idealized impurity [19], or due to added Coulomb interactions [20]. As the temperature increased the differential conductance turned to be only weakly dependent on current and the shot noise reduced. At sufficiently higher temperature, here 120mK, shot noise turned poissonian, with a definite charge $e^*=e/3$ (see also Ref. 21), verifying Laughlin's prediction of fractionally charged quasiparticles (however, flowing in edge channels).

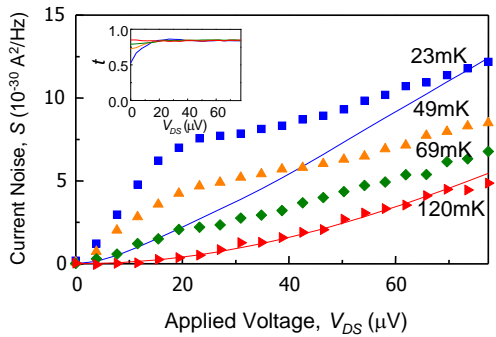

**Fig. 4.** Shot noise and differential conductance at $\nu=1/3$ for different temperatures. Shot noise due to a weak backscattering potential measured at various temperatures: 23, 49, 69 and 120mK. The thin solid lines are the expected shot noise for non-interacting quasiparticles with charge $e/3$ at 23mK (upper line) and 120mK (lower line). Inset: The transmission of the constriction at the same temperatures, with the top most line at 120mK and the lowest one at 23mK.

What happens at an even lower electron temperature, at $T\sim 9$mK? This temperature is very close to the lattice temperature and was achieved after 'cold grounding' of most of the ohmic contacts in the sample (cooling the electron via Wiedemann-Franz- conduction) [22]. As was shown before, a relatively weak backscattering potential at $\nu=1/3$, with a saturated transmission, $t=g/g_{1/3}\sim 0.7$, led to rather strong backscattering in the linear regime (near zero excitation). Moreover, both the voltage and temperature dependences of the differential conductance were positive – as expected from the CLL model ('valley-like' behavior). However, when the QPC split-gate potential was even smaller (in absolute terms), making backscattering weaker, $t=g/g_{1/3}>0.85$, the dependence of the conductance on excitation voltage and temperature reversed sign (see Fig. 5, a 'mound-like' behavior), contradicting the CLL prediction. In these two regimes the charge determined at low excitation voltage is very large (approaching $e$), turning to $e/3$ at higher excitation voltage. Only in the extremely weak backscattering regime (see Fig. 5a, $V_g$=-0.03V), with conductance again independent of excitation voltage the noise was poissonian again, leading to a charge $e^*=e/3$ – very much like at elevated temperatures.

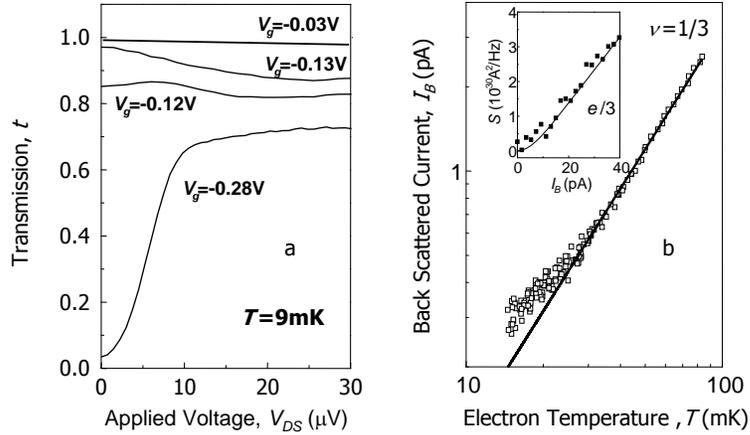

**Fig. 5.** Typical dependence of the transmission coefficient and backscattered current on bias and temperature at $\nu$=1/3. **a.** At different QPC split-gate voltages. When the QPC is very weakly pinched off (split-gate voltage -0.03V, $t$~0.97) the transmission has a very weak (negative) dependence on the applied bias; opposite to CLL behavior. **b.** The backscattered current as a function of electron temperature measured with AC 10μV RMS. The curve can be fitted with a single slope. Inset: The shot noise generated by the weakly pinched constriction as function of the backscattered current, at an electron temperature of 9mK. Noise is classical and quasiparticles charge is $e/3$.

## 5. Shot Noise at $\nu$=2/5 and $\nu$=3/7

What about the charge of quasiparticles in higher CF channels? This measurement is important since it distinguishes between the charge and the conductance. Figure 6a shows, as an example, the linear two terminal conductance as the constriction is being pinched for a bulk-filling factor $\nu_{bulk}=2/5$ (fluctuations are due to instabilities).

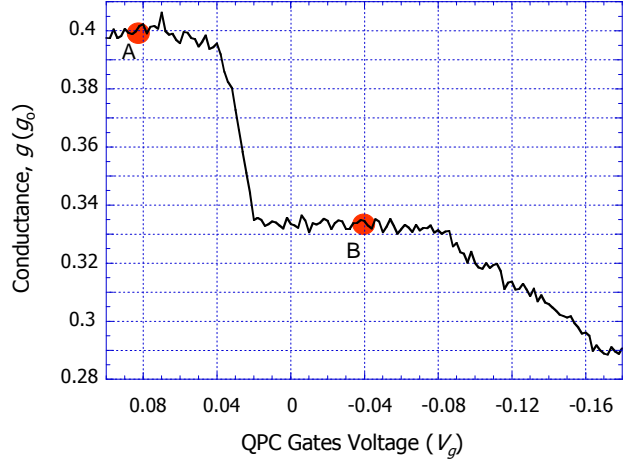

**Fig. 6a.** The conductance as function of the QPC split-gate voltage (affecting the constriction width). Two conductance plateaus, $2e^2/5h$ and $e^2/3h$, are observed.

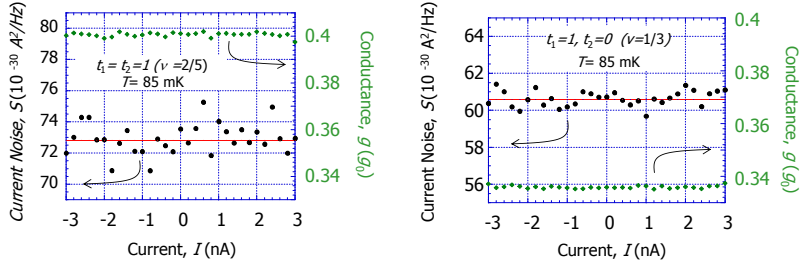

**Fig. 6b.** Noise measured at the two plateaus, at points A and B (Fig. 6a). $t_1$ and $t_2$ refer to the transmission coefficients of the first and second LL of CFs ($\nu=1/3$ and $\nu=2/5$, respectively). The noise is constant and does not depend on the current, accounting only for the thermal and stray noise; hence, the shot noise is zero. The conductance is shown for reference (by + signs).

Figure 6b show the measured noise, as function of the transmitted current in the 2/5 plateau (in the CF language, $t_2=t_1=1$) and in the 1/3 plateau ($t_2=0$, $t_1=1$). As expected, no excess noise (above the thermal and amplifier noises) was measured in each of the plateaus. Since $\langle\delta(i_1)^2\rangle = 0$ and $\langle\delta(i_1+i_2)^2\rangle = \langle\delta(i_1)^2\rangle + \langle\delta(i_2)^2\rangle + 2\langle\delta(i_1 \cdot i_2)\rangle = 0$, the last two terms in the last expression must each be zero independently (unlikely that they cancel each other), leading to the conclusion that the two currents flowing in the two CF channels are independent (last term).

Setting the QPC split-gate voltage to $t\sim0.98$ the temperature dependence of the backscattered current, shown in Fig. 7a, can be fitted with two distinct slopes in $\log(I_B)$ vs $\log(T)$ – with a crossover at $T\sim45$mK. The noise was measured across the full temperature range with two examples plotted in Fig. 7b. Note that with $t_{\mathit{eff}} = \dfrac{t \cdot g_{2/5} - g_{1/3}}{g_{2/5} - g_{1/3}} = 6t - 5$, $t_{\mathit{eff}}$ is substantially smaller than $t$, except for $t$ very close to unity. The two solid lines in Fig. 7b, which agree very well with the data, are the calculated shot noise with charges $q=2e/5$ at $T\sim9$mK and $q=e/5$ at $T\sim82$mK. The charge changes smoothly with temperature over a region of some 20mK (Fig. 7c). While the scattered charge $q=e/5$, which was measured at the *high* temperature region, had been measured already before [23], a charge $q=2e/5$, measured at the lowest temperature, was unexpected. Apparently, at the lowest temperature, in the very weak backscattering regime, one can view the noise as resulting from partitioning of two quasiparticles, $e/5$ each, at each backscatter event (*bunching*).

Similarly, at $\nu=3/7$ the charge approaches $3e/7$ (maximum measured charge was $\sim2.4e/7$) at the lowest available temperature (Fig. 8). Presumably an even lower temperature is needed in order to *bunch* three $e/7$ quasiparticles to $3e/7$ charges. This unexpected bunching might be related to the fractional exchange statistics of the quasiparticles.

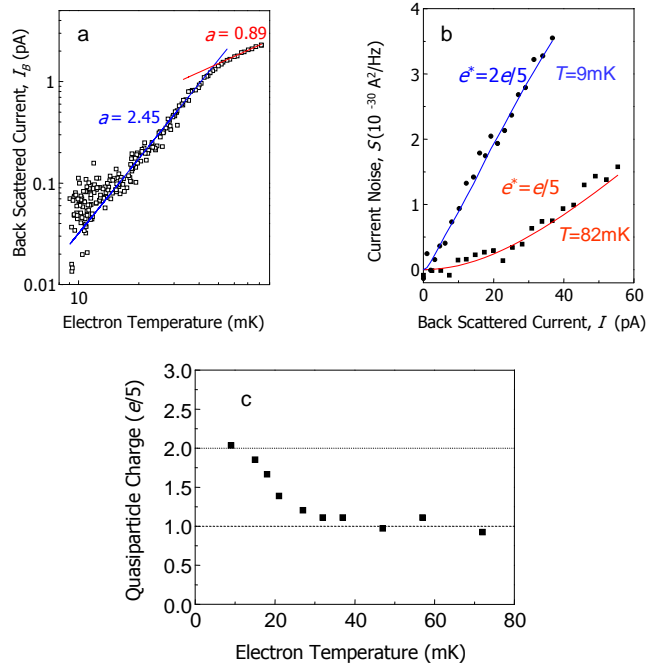

**Fig. 7.** Backscattered current and charge at $\nu=2/5$. **a.** Backscattered current as function of the electron temperature. Two distinct slopes are observed with a transition temperature of about 45mK. **b.** Shot noise at two different temperatures. The backscattered quasiparticle charge is $2e/5$ at 9mK and $e/5$ at 82mK. The constriction was tuned to reflect some 2% of the 2/5 channel at the two temperatures. **c.** The evolution of the quasiparticle charge with temperature.

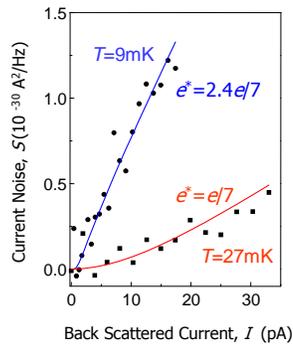

**Fig. 8.** Shot noise at two different temperatures at $\nu=3/7$. The backscattered quasiparticle charge is $\sim 2.4e/7$ at 9mK and $e/7$ at 27mK. The constriction was tuned to reflect the 3/7 channel by some 2%.

## 6. Shot Noise in the Strong Backscattering Limit

In the strong backscattering limit the quasiparticles charge is expected to be different (Eq. 2). As the constriction is being pinched the *I-V* characteristic becomes nonlinear (*g* and *t* depend on current), even at an elevated temperature. Measurements were done at filling factor $\nu=1/3$. To account for the non-linearity, the energy independent Eq. 3 was modified, by approximating the integral over energy in Ref. 14. Moreover, the dependence of the conductance on the current was attributed only to *t*, while the charge $e^*$ was assumed to be constant for a fixed QPC gate voltage. Replacing the integration over energy by a sum over discrete points and substituting *t* in terms of *g* and $e^*$ in Eq. 3, we get [24]:

$$S(I) = 2e^* I \frac{1}{N} \sum_{i=1}^{N} \left(1 - \frac{(g_i/g_0)}{(e_i^*/e)}\right) \left[\coth\left(\frac{e^*V}{2k_BT}\right) - \frac{2k_BT}{e^*V}\right] + 4k_BTg. \quad (4)$$

Here *i* runs over the measured points (*N*) up to current *I*, $g_i$ is the differential conductance at each point, and the term 1-*t* is being replaced by $1-\tilde{t}$, with $\tilde{t}$ the effective transmission coefficient of the *flux* of *quasiparticles* $e^*$ and not the current. The *coth* term, however, rising from integration over the Fermi function, was inserted in 'by hand'. The noise expression now contains a single fitting parameter $e^*$. In other words, for each width of the constriction we find a fitting quasiparticle charge $e^*$ and consequently the channel transmission *t* (via *g*).

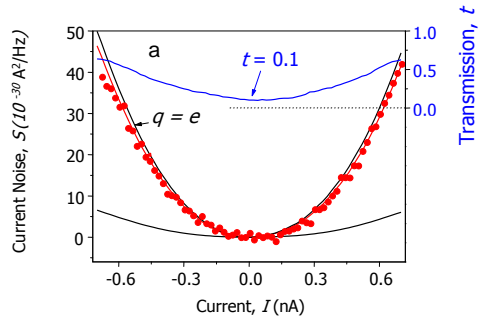

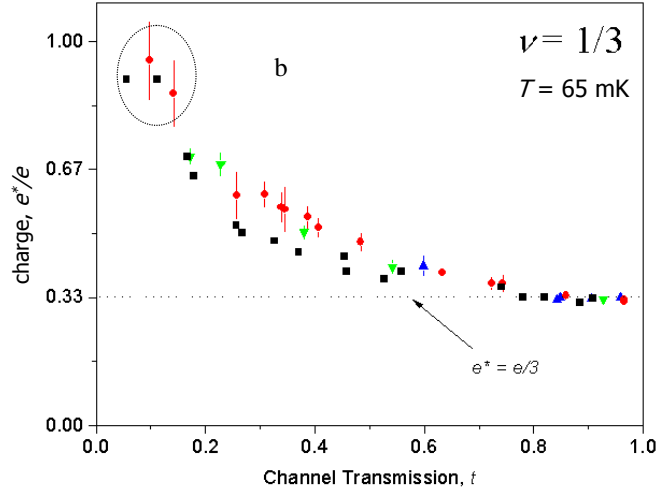

**Fig. 9.** Bunching of *e*/3 quasiparticles at $\nu=1/3$ by a relatively pinched constriction. **a.** Noise and conductance for *t*~0.1, demonstrating bunching of quasiparticles to charge *e*. **b.** The evolution of the charge in three different devices as function of the transmission of the constriction (squares, circles, and triangles stand for different samples).

In Fig. 9a the noise and the fitted charge at $\nu=1/3$ are shown for *t*=0.1 at the lowest applied voltage. The fitted charge is nearly *e*. The dependence of $e^*$ on *t* is summarized in Fig. 9b [24]. Measurements taken on a few samples collapse into one curve. The charge changes smoothly from *e*/3 for weak reflection (large *t*) to around *e* for strong reflection (at *t*~0.1) - as expected.

## 7. Evolution of Charge with Partitioning

We discuss now low temperature (~10mK) shot noise measurements in the non-linear regime, where the transmission is 'mound-like' or 'valley-like'. Shot noise and transmission were measured as function of the excitation voltage $V_{sd}$ and split-gate voltage $V_g$ for bulk filling factors $v=5/2$, $v=7/3$ and $v=1/3$. In the weak and strong backscattering regimes, in all three filling factors, both conductance and spectral density exhibited two distinct regions in the excitation voltage. At low excitation voltage, the conductance, with either 'mound' or 'valley' like behaviors, was accompanied with a large slope in the spectral density, which corresponds to a large charge at all filling factors. At higher excitation voltage a rather 'linear characteristic' and a significantly lower charge were obtained.

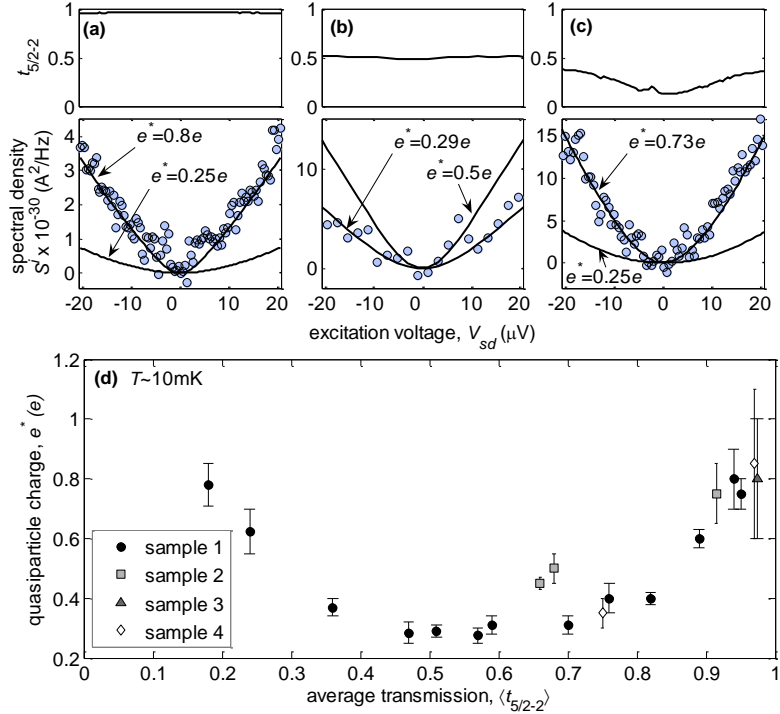

**Fig. 10.** Measurements at $v=5/2$ state of the transmission probability and the shot noise in the small $V_{sd}$ range, measured at $T\sim10$mK. **(a) - (c)** Spectral density for a few transmissions and the predicted spectral densities (using Eq. 1). **(d)** The evolution of quasiparticle charge as a function of the effective (average) transmission probability (of the 5/2 channel with the lower lying channel 2), $\langle t_{5/2-2} \rangle$, measured on four different samples.

Focusing on the low excitation voltage range an interesting dependence of the deduced quasiparticle charge on the average transmission was found. For $v=5/2$, at high, 'mound-like', transmissions ($t\sim0.9$), the charge was substantially higher than the theoretically predicted quasiparticle charge of $e^*=e/4$ (Fig. 10a, and see Ref. 25 for previous measurements and a description of this even-denominator fraction). At intermediate values of the transmission ($t\sim0.4$-$0.9$), with the transmission almost independent of the applied voltage, the charge was very close to $e^*=e/4$ (Fig. 10b). At lower, 'valley-like', transmissions, the charge increased towards $e^*=e$ (Fig. 10c). The latter result is expected since the filling factor within the potential barrier approached $v=2$; enabling thus only backscattering of electrons. Figure 10d summaries the low-excitation-charge evolution as function of the effective transmission $\langle t_{5/2-2} \rangle$, as measured on four different samples. A similar behavior was obtained also for the $v=7/3$ case. The quasiparticle charge evolution as a function of the average effective transmission, $\langle t_{7/3-2} \rangle$, resembles that in the $v=5/2$ state, with $e^*\sim e/3$ obtained only at the intermediate transmission range - where the transmission is nearly voltage independent (not shown).

For the $v=1/3$ state, with measurements performed in the high range of the $t_{1/3}$, with 'mound-like' transmission, and in a somewhat lower transmission which was rather flat. Again, a bimodal quasiparticle charge was observed in the 'mound-like' transmission regime, with $e^*=e$ in the small excitation voltage range and $e^*\sim e/3$ at a higher excitation voltages. For the flat transmission, the charge fitted well $e^*=e/3$ over a wide range of excitation voltage (not shown). The lower transmission range ($t_{1/3}<0.3$), the backscattered charge approached $e$ - as already discussed above [24].

Temperature dependence of the noise measured at high transmissions was studied at bulk filling factor $v=5/2$, as described in Fig. 11. While the longitudinal resistance of the $v=5/2$ state increased weakly with temperature, the resultant backscattering was less than 0.1% at 85mK. As evident from data, the differential transmission coefficient became less sensitive to the applied voltage and the quasiparticle charge reduced. The charge approached $e^*=e/4$ and saturated thereafter as the temperature reached ~75mK and beyond (Fig. 11).

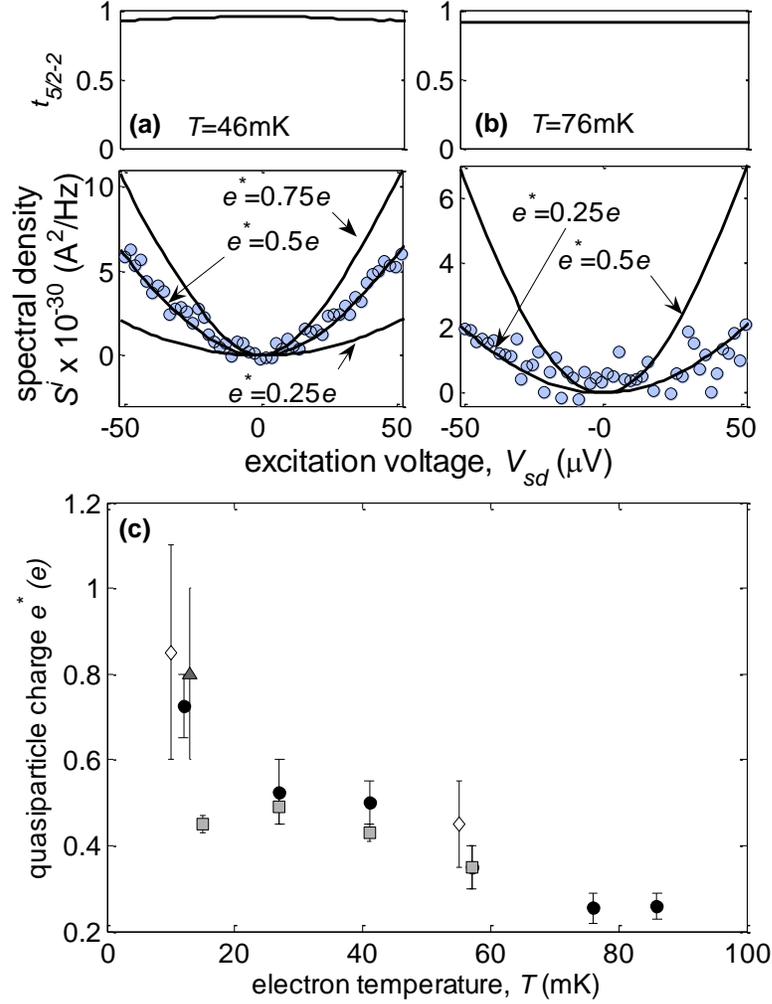

**Fig. 11.** The dependence of the backscattered charge on temperature measured at weak backscattering in the $v=5/2$ state. Two examples of the measured data: **(a)** at $T$=46mK with $e^*$~0.5$e$, and **(b)** at $T$=76mK with $e^*$~$e$/4. **(c)** Evolution of the backscattered charge as a function of temperature. At temperatures higher than ~40mK a significant reduction of the charge is observed.

Our measurements present a rather complex evolution of the determined backscattered quasiparticle charge. It roughly portrays two extreme values of the quasiparticle charge. One small - being close to the 'expected value' - at intermediate backscattering probabilities *or* at higher energies (excitation voltage or temperature), in which the transmission is usually linear. The other one is large, approaching $e^*=e$, for extremely weak backscattering and in the limit of low excitation voltage and temperature, in which the transmission is found to be non linear. This enhancement of the charge can be attributed to either the scattering of an integer multiple of a smaller fundamental charge (*bunching*), or to the fundamental charge being larger than expected. The first assumption is strengthened by the observation of the lower charge at higher excitation voltage or temperature; suggesting that bunching may take place at low energies, and its dissemination into individual fundamental quasiparticle charges at higher energy.

**8. Shot Noise of Composite Edge Channels**
Transport is more complicated if there are one or more counter-propagating edge channels, as is the case of the so called *hole conjugate quantum Hall states*, $v=2/3$, $3/5$, etc. In the case of $v=2/3$, which we describe briefly here, a clean sample devoid of any impurities is expected to support two charged modes: one with conductance of $e^2/h$ - carrying electrons, and a counter-propagating one with conductance $(1/3)e^2/h$, carrying $e/3$ fractional charges [26]. For a smooth edge potential, the two counter-propagating modes will have different momenta (the difference being proportional to the enclosed flux), and hence unlikely to equilibrate. However, in the presence of random inter-channel scattering the momentum of each channel need not be conserved, allowing thus equilibration (and the emergence of a single charge mode) and a universal quantization of Hall conductance $(2/3)e^2/h$. In addition, a *neutral* counter propagating channel (namely, carrying only energy but no charge) is expected to exist [26,27]. Recently, there has been a resurgence of theoretical investigations of this and similar neutral edge channels [28]; ignited by the hypothesis of the much anticipated non-abelian $v=5/2$ fractional state, which is expected to carry a neutral Majorana mode [29].

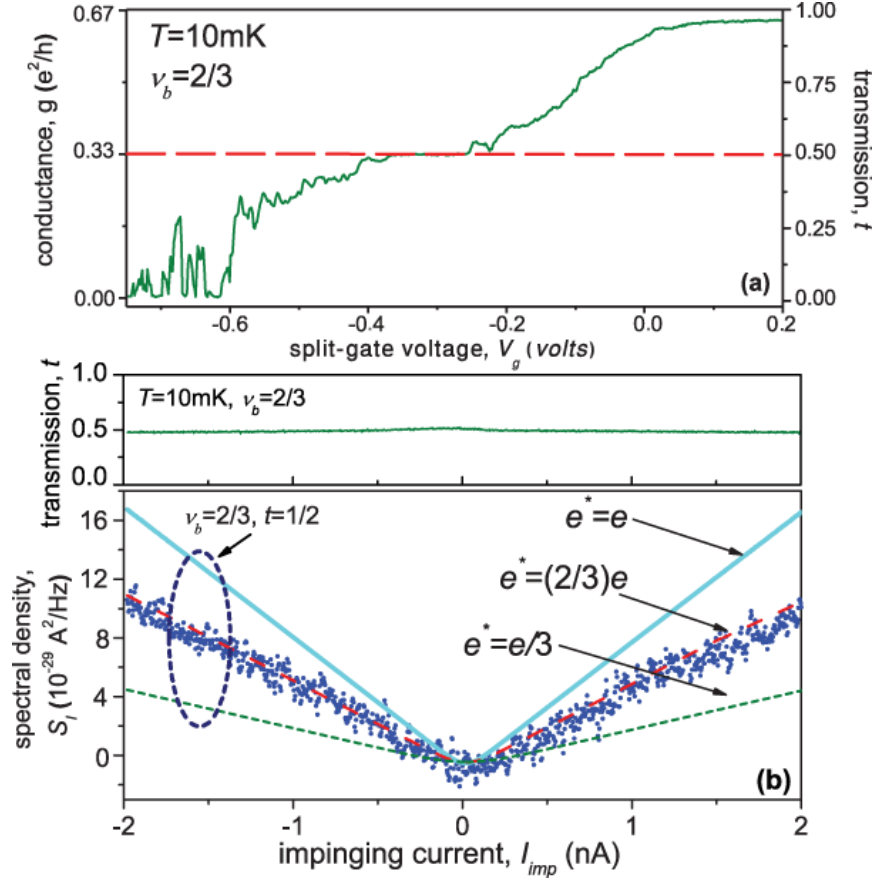

**Fig. 12.** Conductance and spectral density at $v=2/3$ at electron temperature 10mK. (a) Conductance $g$ and transmission $t$ of the constriction as a function of split-gate voltage. Note the appearance of a prominent plateau at $g=e^2/3h$ ($t=1/2$). (b) Upper panel - dependence of the transmission (zero bias $t=1/2$) on injected electron energy. Lower panel - spectral density $S_I$ at this value of transmission. The blue dots are the measured data points. Shown is the expected spectral density for transmission $t=1/2$, temperature $T=10$mK and quasiparticle charge $e^*=e$ (cyan line), $(2/3)e$ (red line) and $e/3$ (olive line).

Figure 12a shows a plot of the transmission as function of the applied split-gate voltage to the QPC, measured on a typical sample. Scanning the gate voltage revealed a prominent plateau at $t=1/2$, suggesting a local filling factor in the constriction $v_C=1/3$. Adopting the assumption that the electron density drops gradually near edges of the constriction, the $v_C=1/3$ plateau seems to confirm that 1/3 edge channel traverses the constriction without backscattering, while the 2/3 edge channel is fully reflected. Under such circumstances the shot noise for $t=1/2$ should be zero. However, surprisingly, this was not the case here. As shown in Fig. 12b, the measured shot noise is *finite*, suggesting a different picture of edge reconstruction in the $v=2/3$ case. Since this behavior repeated itself in all the measured samples, we must adopt the notion of a single chiral-composite-edge channel, with transmission $0<t<1$, determined by the split-gate voltage.

The charge over the entire range of the transmission (from very close to unity to $t\sim0.3$) fits excellently to a value $e^*=(2/3)e$ (not shown). Upon depleting the constriction further the differential transmission became highly current dependent and the spectral density developed two distinct slopes: at the low range of current the quasiparticle charge was $e$ while at higher currents the charge dropped to approximately $(2/3)e$. Since the transmission increases with current (expected in a chiral Luttinger liquid), the charge reverts back to that of the quasiparticle. As the electron temperature increased the charge, measured at $1<t<0.4$, evolved smoothly from $(2/3)e$ to $e/3$. This observation, reminiscent of the behavior of the quasiparticle charge in the 2/5 and 3/7 states [22], may suggest that quasiparticles of charge $e/3$ carry the current in the 2/3 state, however, *bunching* takes place at the low temperature regime.

**9. Shot Noise of Interfering Electrons**
Every interferometer must generate phase dependent shot noise since it splits the incoming electrons into at least two paths, or alternatively, its transmission is in general smaller than one. It is intriguing to measure simultaneously the discrete nature of particles and their interference due to their wave behavior. In a symmetric interferometer, where the incoming beam is split equally between two paths, one can express the transmission $t_{int}=0.5(1+\mu \cos\varphi_{AB})$, with $\mu$ the visibility and $\varphi_{AB}$ the AB phase (the Aharonov Bohm phase), one gets for shot noise:
$$S = 0.5eI_{imp}(1-\mu^2 \cos^2 \phi_{AB}), \quad (5)$$

with $I_{imp}$ the impinging current. Hence, the shot noise contains only the second AB harmonic.

We developed a two-path electron interferometer based on edge channels transport: a Mach-Zehnder electron interferometer [30]. A single edge channel is split and later recombines after enclosing an AB flux. Changing the area of the interferometer, under the quantizing magnetic field, changes the AB phase leading to interference oscillations in the conductance (Fig. 13a). The noise measured under such conditions revealed indeed a second AB harmonic (Fig. 13b).

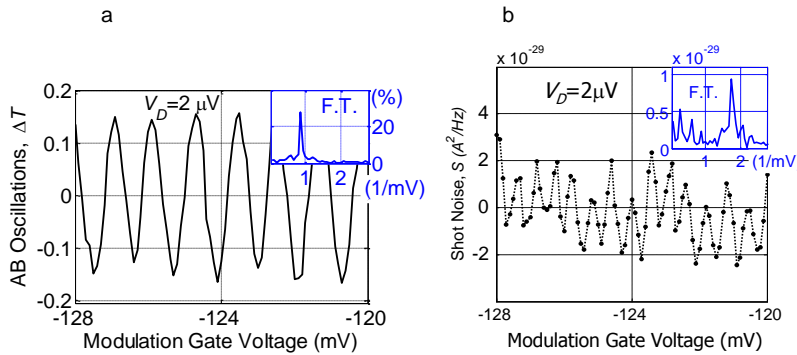

**Fig. 13.** Aharonov-Bohm oscillations of conductance and noise in an electronic Mach-Zehnder interferometer. **a.** The AB oscillations in the conductance as function of the voltage on a gate that changes the area of the interferometer. **b.** Shot noise - second harmonic of the AB oscillations in the conductance.

## Acknowledgements

The work described in this review had been carried out by (in a chronological order), M. Reznikov, R. de-Picciotto, E. Comforti, T. Griffith, and Y-C. Chung, N. Ofek, A. Bid, M. Dolev, and Y. Gross. The material was grown by H. Shtrikman and V. Umansky, and the electron beam lithography done by D. Mahalu. The work enjoyed partial support by the Israeli Science Foundation (ISF), the Minerva Foundation, the German Israeli Foundation (GIF), the German Israeli Project Cooperation (DIP), the US-Israel Bi-National Science Foundation, and the European Research Council under the European Community's

Seventh Framework Program (FP7/2007-2013)/ERC Grant agreement #227716.